\documentstyle[twocolumn,aps,prl,epsfig]{revtex}

\begin{document}
\draft
%\preprint{HEP/123-qed}
\title{Neutron Scattering Investigation of Magnetic Bilayer Correlations
in La$_{1.2}$Sr$_{1.8}$Mn$_2$O$_7$}
\author{R. Osborn$^1$, S. Rosenkranz$^1$, D.N. Argyriou$^2$, L. 
Vasiliu-Doloc$^{3*}$, J. W. Lynn$^{3*}$, S.K. Sinha$^4$, J.F. Mitchell$^1$,
K.E. Gray$^1$, S.D. Bader$^1$}
\address{$^{1}$Materials Science Division, Argonne National Laboratory, Argonne,
Illinois 60439}
\address{$^2$Manual Lujan, Jr. Neutron Scattering Science Center, Los Alamos 
National Laboratory, Los Alamos, New Mexico 87545}
\address{$^3$NIST Center for Neutron Research, National Institute of Science and
Technology, Gaithersburg, Maryland 20899}
\address{$^4$Advanced Photon Source, Argonne National Laboratory, Argonne, 
Illinois 60439}
\date{\today}
\maketitle
\begin{abstract}
Neutron scattering investigations of the paramagnetic correlations in the 
layered manganite La$_{1.2}$Sr$_{1.8}$Mn$_{2}$O$_{7}$, which exhibits colossal 
magnetoresistance above the Curie transition at T$_C$ = 112 K, show that spins 
in neighboring layers within each bilayer are strongly canted at an average 
angle that is dependent on both the magnetic field and temperature, as predicted
by de Gennes.  The in-plane correlation length does not diverge at T$_C$,
although the magnetic Bragg intensity obeys critical scaling below T$_C$, with
the same temperature dependence as the zero-field electrical conductance.
\end{abstract}

%\newpage

Naturally layered manganites have proved to be fruitful systems for 
understanding the mechanism of colossal magnetoresistance (CMR) and have become 
the focus of many  recent investigations into this phenomenon 
\cite{moritomo,kimura,perring,mitchell,argyriou,battle}. The reduced 
dimensionality increases the magnitude of the CMR, although at the cost of 
reducing the ferromagnetic transition temperature to about 100 K 
\cite{moritomo}. However, the extended temperature range over which
ferromagnetic correlations are significant and their strong anisotropy allow
a detailed examination of the link between local spin correlations and the 
resulting magnetotransport. The extra freedom afforded the crystal chemist by 
the layered structure may also lead to better-optimized materials for
industrial applications.

The majority of experiments have been performed on the Ruddlesdon-Popper phases
with the general formula (La$_{1-x}$Sr$_x$MnO$_3$)$_n$SrO, which comprise $n$
layers of corner-shared MnO$_6$ octahedra separated by (La,Sr)O blocking layers.
In particular, the two-layer compounds have revealed a rich variety of properties which are strongly dependent on $x$ \cite{kimura}. Although most 
investigations of three-dimensional CMR compounds have concentrated on the 
possible role of electron-lattice interactions \cite{millis,roder,teresa}, of
equal importance in the two-dimensional compounds has been the influence of
antiferromagnetic interactions competing with the ferromagnetic double-exchange.
Perring {\it et al} \cite{perring} have reported evidence of weak 
antiferromagnetic correlations within the planes above T$_C$ that are 
fluctuating rapidly and are believed to coexist with the ferromagnetic
correlations. On the other hand, Argyriou {\it et al} \cite{argyriou} have
inferred the existence of a canting of the ordered moments below T$_C$ from the
change in sign of Mn-O bond compressibilities at the transition. This aspect
of the problem is only now receiving the theoretical attention it deserves,
even though de Gennes first considered it nearly forty years ago
\cite{degennes,mishra,golosov}.

The work reported here is part of a general study linking the magnetic and 
transport properties of naturally layered manganites directly to the underlying
magnetic correlations measured by neutron diffraction and spectroscopy. We find
evidence that, although the magnetic correlations are predominantly 
ferromagnetic within the two-dimensional planes, there is much weaker 
ferromagnetic correlation between spins in neighboring layers within each
bilayer. This observation is consistent with a canting of the spins in
neighboring layers with a cant angle that is dependent on both magnetic field
and temperature, becoming smaller as the temperature approaches T$_C$. The
correlation is not confined to short-lived clusters, as were the 
antiferromagnetic fluctuations observed by Perring {\it et al} \cite{perring}, 
but involves the critical fluctuations that lead to ferromagnetic ordering. 
Our results imply that there is a delicate balance between competing double 
exchange and superexchange interactions in these compounds. In the critical 
regime below T$_C$, the link between magnetic ordering and electrical transport
is strikingly evident, with both displaying similar two-dimensional exponents.

An x = 0.4 single crystal, which we will refer to as the 40\%-doped compound,
La$_{1.2}$Sr$_{1.8}$Mn$_2$O$_7$, weighing 150 mg was grown in a double-mirror,
floating zone image furnace. Bulk measurements show that the sample has a Curie
transition at about T$_C$ = 112 K, which also corresponds to the change from
insulating to metallic transport both parallel and perpendicular to the MnO$_2$
planes. The lattice parameters are $a = 3.862$ $\AA$ and $c = 20.032$ $\AA$
at 125 K in good agreement with earlier measurements for this composition
\cite{mitchell}. We saw evidence of a small crystallite in the sample with 
slightly smaler values of $a$ and $c$ and a T$_C$ value of 114 K, which could 
therefore have slightly different dopant level. The only time that this 
presented a significant problem was when measuring the order parameter.
However, improved collimation ensured that we were only measuring scattering
from one crystallite. Diffuse neutron scattering studies were performed on the 
Single Crystal Diffractometer (SCD) at Los Alamos where a broad region of reciprocal 
space could be measured in one scan using a pulsed white beam. These were 
followed up by measurements using the triple-axis spectrometers BT2 and BT9 at
Gaithersburg. For the measurements at BT9, we employed a superconducting
solenoid to provide fields to 7 T applied in the $ab$ plane. Using incident 
wavevectors of k$_i$ = 2.57 and 2.662 $\AA^{-1}$ without energy analysis, we performed scans in the [0$k$0] scattering plane {\it i.e.} with the [$h$00] and
[00$l$] reciprocal lattice vectors as orthogonal axes within the plane. The 
conductance was measured using a six-probe technique suitable for materials with
highly anisotropic conductivities, such as the cuprate superconductors 
\cite{busch}.

The SCD measurements, illustrated in Fig. 1, show that the scattering above 
T$_C$ consists of rods parallel to the $c$-axis, {\it i.e.} parallel to [00$l$],
at all integer $h$ including $h = 0$. The rods are strongly 
temperature-dependent, becoming narrower as T $\to$ T$_C$, consistent with 
magnetic scattering from predominantly ferromagnetic correlations which have 
much longer correlation lengths within the MnO$_2$ planes than perpendicular
to them. Below T$_C$, the scattering results from inelastic spin wave 
excitations and so becomes much weaker with decreasing temperature. We also
observe a weaker second component, peaked at $l$ = 0, with a longer in-plane 
correlation length. Since this is nearly temperature-independent near T$_C$
and has both magnetic and nuclear components, we believe that it arises from a
low density of multilayer intergrowths, as identified by Potter {\it et al}
\cite{potter}, and have excluded it from our analysis.

In the triple-axis experiments, scans were performed both parallel and 
perpendicular to the rods (00$l$) and (10$l$). The magnetic neutron scattering
cross section is proportional to
\begin{eqnarray}
S({\bf Q})&=&\sum_{\alpha \beta} S^{\alpha \beta} ({\bf Q})\nonumber \\
&=&\sum_{\alpha \beta} \left ( \delta_{\alpha \beta} - 
\hat{Q}_{\alpha} \hat{Q}_{\beta}\right ) 
\langle S_{\alpha} ({\bf Q}) S_{\beta} (-{\bf Q}) \rangle
\end{eqnarray}

\begin{center}
with $\displaystyle S_{\alpha} ({\bf Q}) = \frac{1}{\sqrt N} \sum_{i=1}^{N} S_{\alpha, i} 
exp(i {\bf Q} \cdot {\bf r}_i),$
\end{center}
where {\bf Q} is the wavevector transfer, and $S_{\alpha, i}$ and 
${\bf r}_i$ are the magnetic moments and atomic coordinates, respectively,
of the Mn ions ($\alpha, \beta$ = x,y,z; the z-direction is parallel to the
crystal $c$-axis). Because of the orientation factor in the paranthesis of
Eq. 1, scans are only sensitive to magnetic fluctuations perpendicular to 
{\bf Q}. Therefore, scans along {\bf Q} = [00$l$] will only measure in-plane 
correlations, $S^{xx} ({\bf Q}) + S^{yy} ({\bf Q})$, while scans along [$h$0$l$]
will be dominated by $S^{yy} ({\bf Q}) + S^{zz} ({\bf Q})$, for small values of 
$l$. By combining the results of scans along [0.05 0 $l$] and [0.95 0 $l$], we 
have been able to separate $S^{xx,yy} ({\bf Q})$ from $S^{zz} ({\bf Q})$.

If the only significant correlations are between Mn spins within a bilayer,
{\it i.e.} at ${\bf r}_i = \pm z{\bf c}$, then the rod scattering will be modulated as a function of $l$ such that 
\begin{eqnarray}
S({\bf Q})&=&S^2 [ \frac{1}{2} \left ( 1 + \hat{Q}_z^2 \right )
\langle cos^2 \gamma \rangle ( 1 + R \langle cos \theta \rangle cos 4\pi zl )
\nonumber \\
& &+ \left ( 1 - \hat{Q}_z^2 \right ) \langle sin^2 \gamma \rangle ],
\end{eqnarray}
where $\theta$ is the in-plane angle between spins in neighboring layers 
within the bilayer, $\gamma$ is the angle of the spin with respect to the 
planes, and R = $\langle cos \gamma \rangle^2 / \langle cos^2 \gamma \rangle$
assuming that the $z$-axis spin components are uncorrelated 
($8/\pi ^2 \leq R \leq 1$). Since S({\bf Q}) is determined by instantaneous (t =
0) spin correlations, the angular brackets represent ensemble averages over the
crystal. In the 40\% compound, $z$ is 0.0964 so that a ferromagnetic modulation 
would peak at $l$ = 0 and fall to zero at $l$ = 2.59.

Short-range spin correlations within the plane produce a Lorentzian broadening of
the rods with the half-width equal to the inverse correlation length. This is seen 
in Fig. 2(a), which shows a scan perpendicular to the [10$l$] rod at $l$ = 1.833, 
from which we can estimate the ferromagnetic in-plane correlation length to be 9.7
$\AA$ at 125 K. Scans at other values of $l$ show that the correlation length is 
constant along the rod as expected.
  
The most striking observation is that the modulation along the rod, shown in 
Fig. 2(b) is very weak even close to T$_C$, indicating that 
$\langle cos \theta \rangle \ll$ 1. From fits to Eq. 2, we have determined that 
$\langle cos \theta \rangle \simeq$ 0.06 at 125 K in zero field, whereas,
if the ferromagnetic correlations between the spins at $\pm z{\bf c}$ were as 
strong as those within the plane, then $\langle cos \theta \rangle$ would be 
approximately 0.67.  Since we only measure the average value of $cos \theta$, and 
not its distribution, there are two reasonable interpretations of this observation.\\
The first is that the average value of $\theta$ is zero but that the correlations 
between the neighboring planes are much weaker than within the plane.  In the double 
exchange mechanism, the mobile electrons lower their kinetic energy by polarizing 
the localized Mn spins, so the free energy gain from delocalizing the electrons would
be greater within the planes, where a large number of spins can participate in the 
ferromagnetic cluster, than between the neighboring planes, where only two sites are 
involved.  Nevertheless, the nearest-neighbor Mn-Mn distance between the planes is 
the same as within the plane, so it seems unlikely that the interplanar coupling 
would be more than an order of magnitude weaker than the intraplanar coupling, as 
required to explain the disparity in correlations. \\ 
The second interpretation is that there is a canting of the spins in neighboring 
planes, {\it i.e.} the average value of $\theta$ is non-zero.  Measurements of 
spinwave energies in LaMnO$_3$ \cite{hirota,moussa} have provided evidence of 
competing ferro- and antiferromagnetic exchange interactions of the usual Heisenberg 
form in a related structure. However, nearest-neighbor Heisenberg exchange, whose 
energy is proportional to $S^2 cos \theta$, will only produce collinear ferro- or 
antiferromagnetic ordering, except when strong magnetocrystalline anisotropy tilts 
the spins away from symmetry directions. In La$_{1.2}$Sr$_{1.8}$Mn$_2$O$_7$, there 
is no buckling of the MnO$_2$ planes to induce such tilts \cite{mitchell}.  On the 
other hand, the energy of double exchange interactions for a pair of spins depends 
on $cos \theta/2$ \cite{anderson} making non-collinear spin correlations, {\it i.e.} 
with a minimum free energy at $\theta \neq 0$ or $\pi$, in zero field a possibility 
when double exchange and superexchange are competing, as first pointed out by de 
Gennes \cite{degennes}.  The observed modulation gives an average cant angle of 
$(86.6 \pm 1.5)^{\circ}$ at 125 K in zero field which suggests that the competing 
interactions are of the same order of magnitude in this compound.

We have repeated the scans with a magnetic field applied vertically {\it i.e.} 
parallel to [0$k$0]. As expected, the modulation becomes stronger with field [see 
Fig. 2(b)], corresponding to a reduction of the cant angle at 125 K to 
$(74.1 \pm 2.1)^{\circ}$ at H = 1 T, although there is very little change at H = 0.5 T.
Increasing the field still further to 2 T reduces the cant angle to $(53 \pm
6)^{\circ}$, but results in a strong decrease of the scattering intensity because 
the spin fluctuations have been reduced by the growth of three-dimensional (3D) 
magnetic order.
 
When the temperature is reduced to T$_C$, the modulation increases so that the 
zero-field and 0.5 T  data at 112 K are very similar to the 1 T and 2 T data, 
respectively, at 125 K.  Since the correlations between different bilayers have also 
built up considerably at this temperature, with increases in scattering around the 
3D Bragg points such as at (101), it appears that the additional interplanar exchange
coupling favors a smaller cant angle. Below T$_C$, Argyriou {\it et al} 
\cite{argyriou} have shown that the ferromagnetic phase is probably also canted but 
the precise angle is not known. 

We have also determined the temperature dependence of the in-plane correlation 
lengths. In order to optimize the energy integration of the scans, we have performed 
them with the scattered wavevector, $k_f$, parallel to the rods.  Surprisingly, the 
correlation length does not appear to diverge at T$_C$ only reaching a value of 
12.1 $\AA$ (Fig. 3).  Although this means that we are never in the critical regime, 
power-law scaling of the correlation length is consistent with our results, with an 
exponent $\nu$ that depends strongly on the assumed value of the 2D critical
temperature, T$_C^{2D}$, {\it i.e.} the temperature at which 2D ordering would occur 
in the absence of 3D interactions. A mean field exponent of $\nu$ = 0.5 gives 
T$_C^{2D}$ = 98.2 K while, at the other extreme, the 2D-Ising exponent $\nu$ = 1 
gives T$_C^{2D}$ = 63.0 K.  
 
The lack of divergence of the 2D-correlation length at T$_C$ may be evidence that 
the phase transition is actually weakly first-order.  For example, either 
electron-lattice interactions may play a role in driving this transition, since 
lattice parameter anomalies are also observed in this temperature range 
\cite{mitchell}, or the ordering may be unconventional, {\it e.g.} consider the 
small-to-large polaron transition proposed by R\"{o}der {\it et al} \cite{roder}.  
However, there are also signs of a build-up of 3D correlations from 120 to 112 K, so 
it may be that there is a crossover to 3D critical scaling very close to T$_C$.  
Since recent spinwave measurements indicate that the exchange coupling within the 
bilayer is two orders of magnitude greater than between bilayers \cite{rosenkranz}, 
such a 3D crossover would be expected to occur within only $\sim$1 K of T$_C$.

Below T$_C$, we observe scaling of the magnetic order parameter with no clear 
evidence that the transition is interrupted by a first-order transition 
\cite{1stord}. Scans through the (002) Bragg peak, which has extremely weak nuclear 
intensity, show that the intensity, which is proportional to M$^2$, scales as 
$t^{2\beta}$ where $t$ = (T$_C$ - T) / T$_C$ from 100 to 111 K, with T$_C$ = 
$(111.7 \pm 0.2)$ K and $\beta = 0.13 \pm 0.01$. The small value for $\beta$ indicates 
that the fluctuations below T$_C$ are still strongly two-dimensional ({\it e.g.}
$\beta$ = 0.125 in the 2D Ising model). Figure 4 shows that there is a strong 
fluctuation tail that extends to $\sim$3\% above the scaling value of T$_C$.  
Superimposed on the neutron data in Fig. 4 are in-plane conductance results 
on a single crystal (grown in the same way as the neutron sample) which has T$_C$ = 
113.3 K and $\beta = 0.17 \pm 0.01$ determined between 90 and 110 K.  When scaled to 
the slightly different T$_C$ values, the coincidence of the two sets of data, even 
into the fluctuation regime above T$_C$, is manifest. This shows that in the 
critical regime, the electrical conductance has the same temperature dependence as 
the squared zero-field magnetization, a direct correlation which, to our knowledge, 
has not been predicted \cite{furukawa,okabe}. If this connection is universal, it 
will constrain theoretical approaches to the spontaneous magnetic state below T$_C$.

In conclusion, we have observed evidence that Mn spins in neighboring layers within 
each bilayer of the naturally-layered CMR manganite La$_{1.2}$Sr$_{1.8}$Mn$_2$O$_7$ 
are strongly canted above T$_C$ at an angle that is both temperature and field 
dependent. It is possible to identify this canting because of the reduced 
dimensionality of the magnetic correlations, so if it also occurs in the regular 
perovskites, it would be more difficult to observe. The growth of 3D correlations 
close to T$_C$ produces a reduction in the average cant angle although the in-plane 
correlation length never fully diverges, possibly indicating that the transition is 
weakly first-order. The direct link between magnetic order and electronic transport 
is clearly seen in the 2D critical scaling just below T$_C$.  

We have benefited from extensive discussions with D. I. Golosov, M. R. Norman and K. 
Levin.  The work was supported by the U.S. Department of Energy, Basic Energy 
Sciences, Division of Materials Science, under Contract Nos. W-31-109-ENG-38 and 
W-7405-ENG-36. Research at the University of Maryland was supported by the NSF-MRSEC,
DMR 96-32521.

$^*$Also at the University of Maryland, College Park, MD 20742

\newpage

\begin{center}
{\large FIGURE CAPTIONS }
\end{center}

FIG. 1. Diffuse neutron scattering above T$_C$ (at 130 K) in the ($0k0$) plane
showing the rod of magnetic scattering along the [$h0l$] direction. \\

FIG. 2. (a) Diffuse neutron scattering at 125 K along the {\bf Q} = [$h$ 0 
1.833] direction. The solid line is a fit to a Lorentzian lineshape convolved
with the instrumental resolution. The half-width, $\kappa = 0.102$ $\AA^{-1}$,
is the inverse correlation length within the two-dimensional planes.
(b) Diffuse scattering at 125 K along the {\bf Q} = [0.95 0 $l$] direction
with an applied field of 0 T (filled circles), 1T (open circles) and 2 T 
(filled squares). The solid lines are fits to Eq. 2 with 
$\theta = 86.6^{\circ}$, $74.1^{\circ}$ and $53^{\circ}$, respectively. \\

FIG. 3. In-plane correlation length {\it vs} temperature. The solid line is a 
power-law fit to the data with T$_C^{2D} = (75 \pm 11$) K and exponent 
$\nu = 0.84 \pm 0.15$. The dashed and chain lines show fits using the 2D Ising 
and mean field values of $\nu = 1$ and 0.5, respectively. The inset shows the 
inverse correlation length similarly fitted. The arrows show the value of T$_C$ 
= 111.7 K determined from scaling below the transition. \\

FIG. 4. Intensity of (002) Bragg peak {\it vs} reduced temperature 
(T$_C$ - T)/T$_C$. The dashed line is a power-law fit from 100 K to 111 K with 
T$_C = (111.7 \pm 0.2$) K and $\beta = 0.13 \pm 0.01$. The solid line is the 
in-plane electrical conductance measured on a sample of the same composition
with T$_C$ = 113.3 K. The inset shows the same measurements over an expanded
temperature range.

\newpage
\clearpage
\widetext
\vspace*{0cm} 
\hspace*{0cm} \epsfig{file=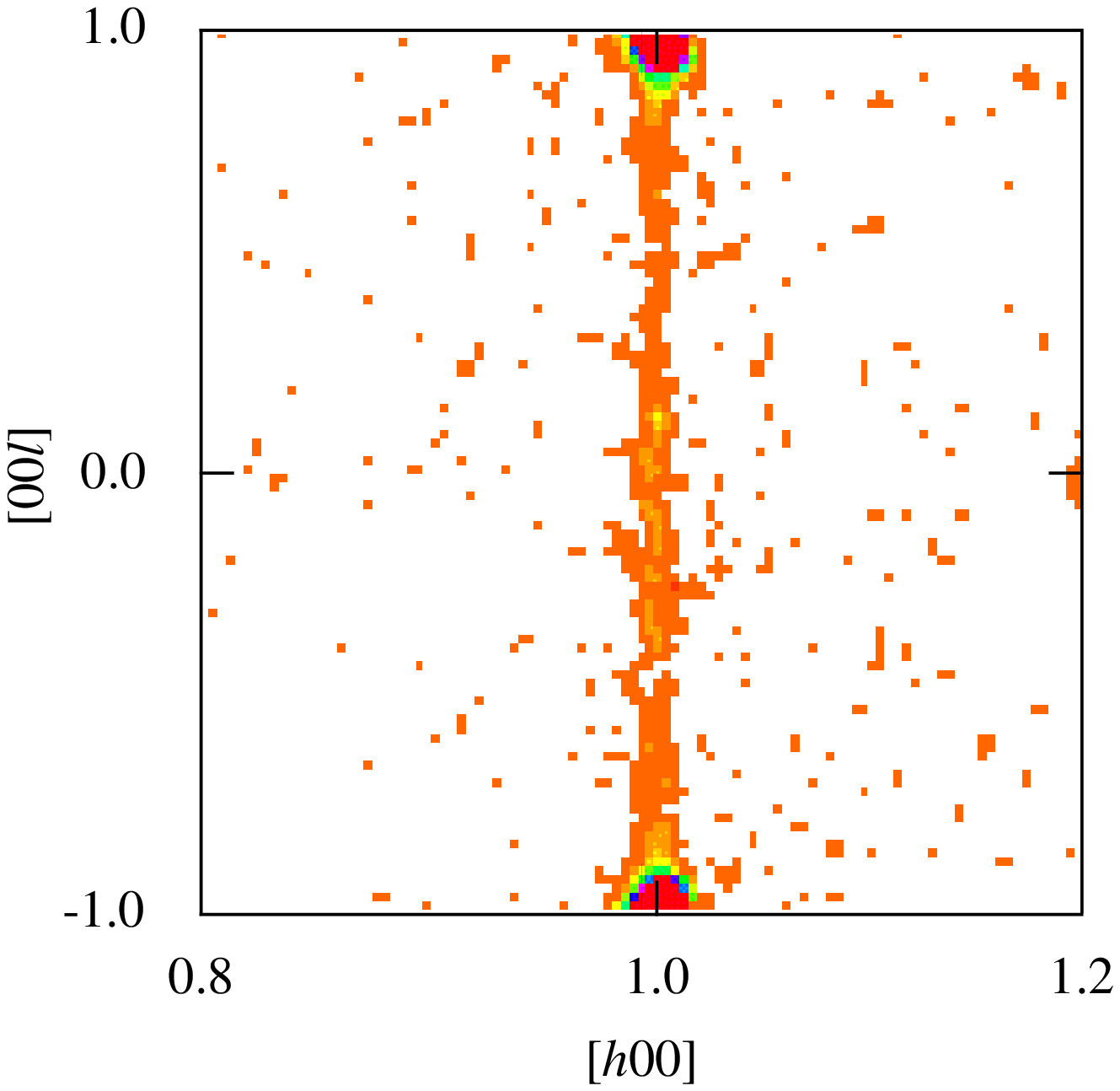,width=17truecm}

\vspace{-1cm}
\large
\hspace{13cm} FIG. 1

\newpage
\clearpage
\vspace*{1.5cm}
\hspace*{2.0cm} \epsfig{file=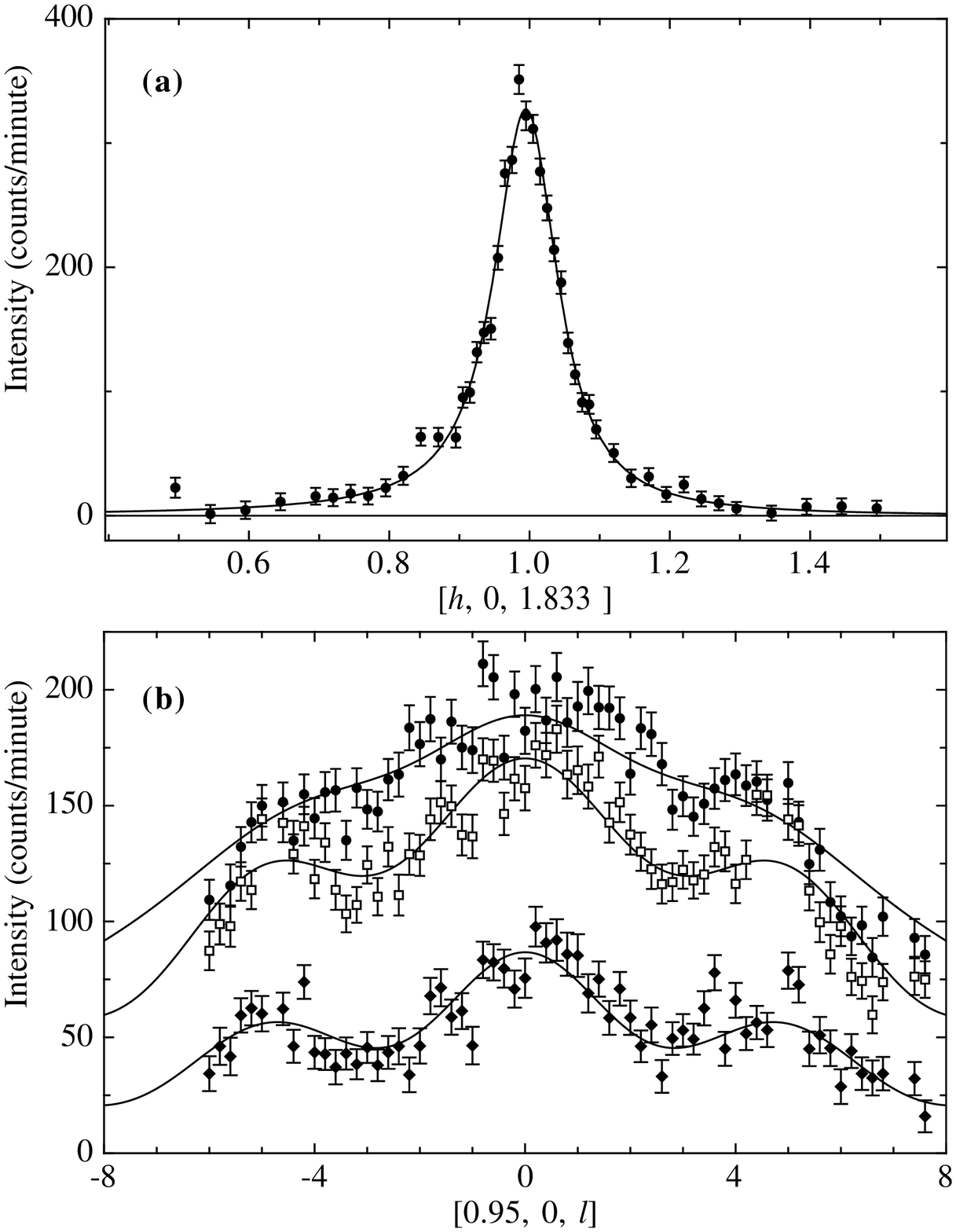,width=15truecm}

\vspace{0cm}
\large
\hspace{13cm} FIG. 2

\newpage
\clearpage
\vspace*{2.0cm}
\hspace*{1.0cm} \epsfig{file=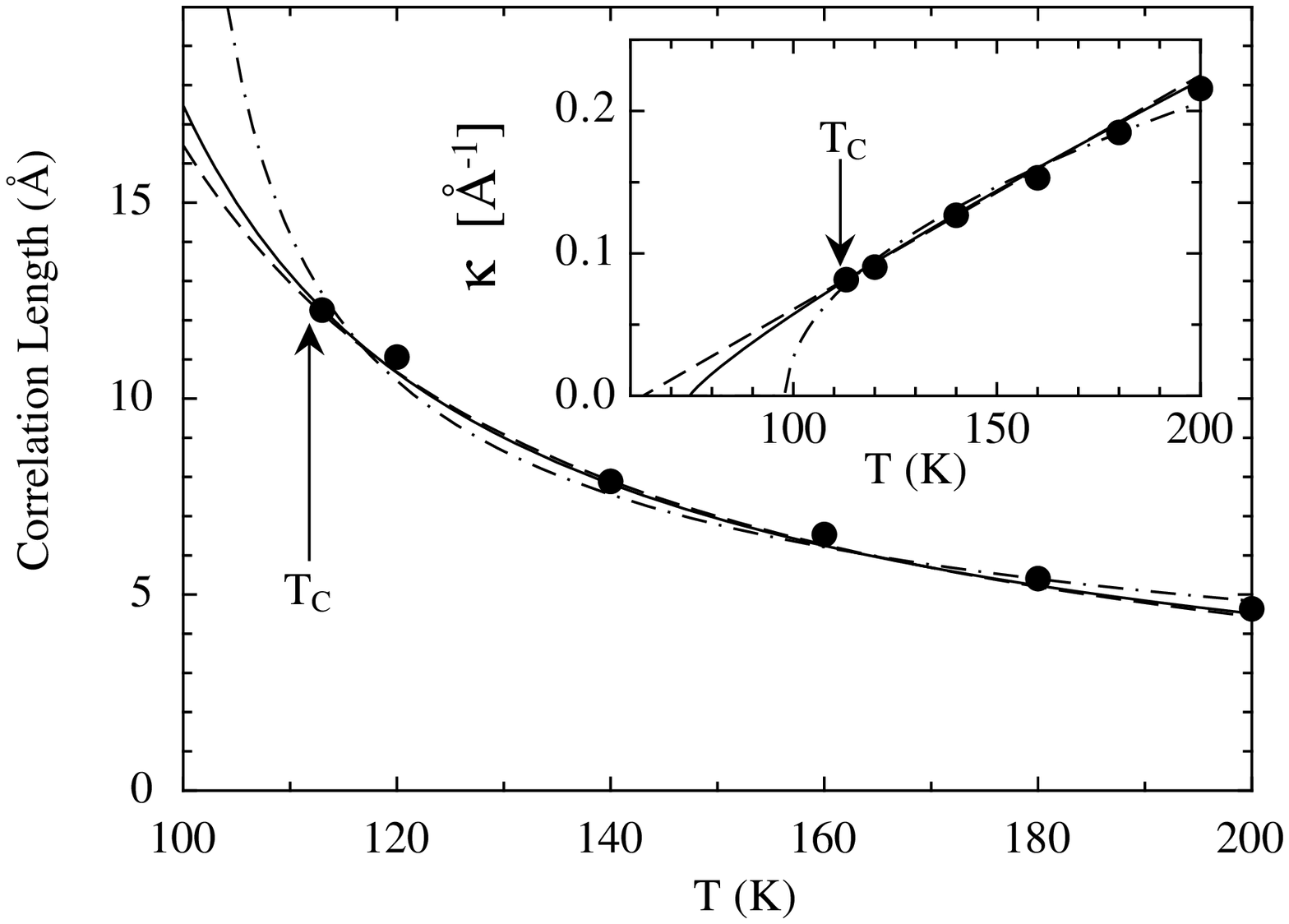,width=16truecm}

\vspace*{-2.0cm}
\hspace*{13cm} FIG. 3

\newpage
\clearpage
\vspace*{0cm}
\begin{center}
\hspace*{1cm} \epsfig{file=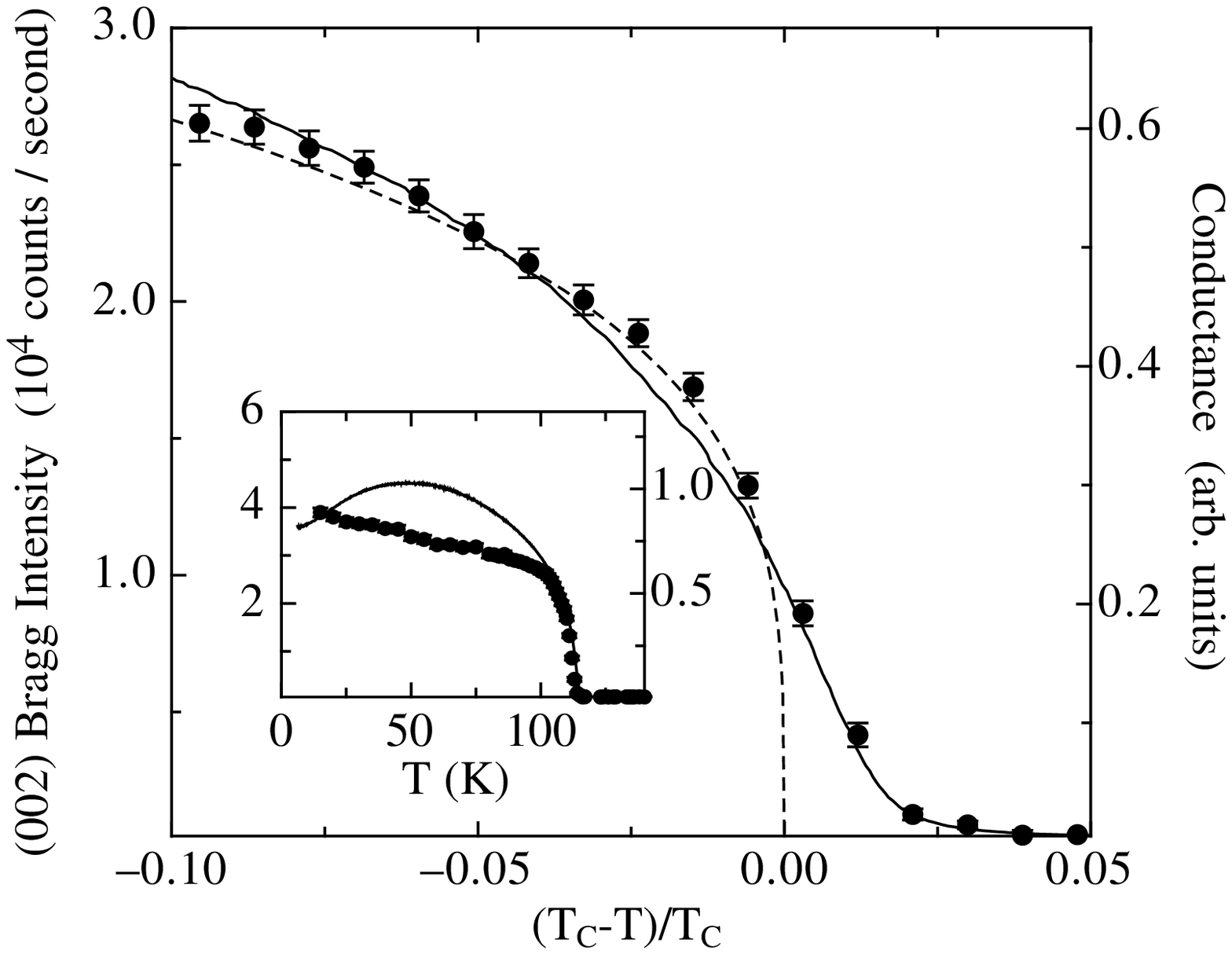,width=17truecm}
\end{center}

\vspace*{-3.0cm}
\large
\hspace{13cm} FIG. 4


\begin{references}
\bibitem{1stord} The absence of any hysteretic behavior also indicates that, if 
the transition is first-order, it is only weakly so.
\end{references}
\end{document}